\documentclass[journal]{IEEEtran}
\usepackage[colorlinks,linkcolor=magenta,anchorcolor=magenta,citecolor=magenta,bookmarks=true]{hyperref}  
\usepackage[T1]{fontenc}
\ifCLASSINFOpdf
\usepackage[pdftex]{graphicx} 
\DeclareGraphicsExtensions{.eps,.pdf,.jpeg,.png}
\else
\usepackage[dvips]{graphicx}
\fi
\usepackage[compress,nospace]{cite}
\usepackage[cmex10]{amsmath}
\usepackage{epstopdf,amsthm,stfloats,siunitx,amssymb,wasysym,algorithm,algorithmic,array,url, color,subfigure} 
\usepackage{footnote}
\usepackage{steinmetz}
\makesavenoteenv{tabular}
\newtheorem{theorem}{Proposition}

\usepackage{soul}

\begin{document}
%
% paper title
% Titles are generally capitalized except for words such as a, an, and, as,
% at, but, by, for, in, nor, of, on, or, the, to and up, which are usually
% not capitalized unless they are the first or last word of the title.
% Linebreaks \\ can be used within to get better formatting as desired.
% Do not put math or special symbols in the title.
\title{DNN-DANM: A High-Accuracy Two-Dimensional DOA Estimation Method Using Practical RIS}
%
%
% author names and IEEE memberships
% note positions of commas and nonbreaking spaces ( ~ ) LaTeX will not break
% a structure at a ~ so this keeps an author's name from being broken across
% two lines.
% use \thanks{} to gain access to the first footnote area
% a separate \thanks must be used for each paragraph as LaTeX2e's \thanks
% was not built to handle multiple paragraphs
%

\author{Zhimin~Chen,~\IEEEmembership{Member,~IEEE,}  Peng~Chen,~\IEEEmembership{Senior Member,~IEEE,} Le~Zheng,~\IEEEmembership{Senior Member,~IEEE,} Yudong~Zhang,~\IEEEmembership{Senior Member,~IEEE} 
\thanks{Copyright (c) 2015 IEEE. Personal use of this material is permitted. However, permission to use this material for any other purposes must be obtained from the IEEE by sending a request to pubs-permissions@ieee.org.}
\thanks{This work was supported in part by the Natural Science Foundation of Shanghai under Grant 22ZR1425200, the Natural Science Foundation for Excellent Young Scholars of Jiangsu Province under Grant BK20220128, the Open Fund of ISN State Key Lab under Grant ISN24-04,
		the National Key R\&D Program of China under Grant 2019YFE0120700, the National Natural Science Foundation of China under Grant 61801112, and the Shanghai Science and Technology Plan Project under Grant 21010501000. \textit{(Corresponding author: Peng Chen)}}
\thanks{Z.~Chen is with the School of Electronic and Information, Shanghai Dianji University, Shanghai 201306, China, with State Key Laboratory of Integrated Services Networks, Xidian University, Xi'an 710071, China, and also with the Department of Electronic and Information Engineering, The Hong Kong Polytechnic University, Hong Kong (e-mail: chenzm@sdju.edu.cn).}
\thanks{P.~Chen is with the State Key Laboratory of Millimeter Waves, Southeast University, Nanjing 210096, China, and also with State Key Laboratory of Integrated Services Networks, Xidian University, Xi'an 710071, China (e-mail: chenpengseu@seu.edu.cn).}
\thanks{L. Zheng is with the Radar Research Laboratory, School of Information and Electronics, Beijing Institute of Technology, Beijing 100081, China, and is also with the Chongqing Innovation Center, Beijing Institute of Technology, Chongqing 401120, China (e-mail: le.zheng.cn@gmail.com).}
\thanks{Y. Zhang is with the School of Computing and Mathematical Sciences, University of Leicester, Leicester, LE1 7RH, UK (e-mail: yudongzhang@ieee.org).} 
}

\markboth{IEEE Transactions on Vehicular Technology}%
{Shell \MakeLowercase{\textit{et al.}}: Bare Demo of IEEEtran.cls for IEEE Journals}
 
\maketitle
 
\begin{abstract}
Reconfigurable intelligent surface (RIS)  or intelligent reflecting surface (IRS) has been an attractive technology for future wireless communication and sensing systems. However, in the practical RIS, the mutual coupling effect among RIS elements, the reflection phase shift, and amplitude errors will degrade the RIS performance significantly. This paper investigates the two-dimensional direction-of-arrival (DOA) estimation problem in the scenario using a practical RIS. After formulating the system model with the mutual coupling effect and the reflection phase/amplitude errors of the RIS, a novel DNN-DANM method is proposed for the DOA estimation by combining the deep neural network (DNN) and the decoupling atomic norm minimization (DANM). The DNN step reconstructs the received signal from the one with RIS impairments, and the DANM step exploits the signal sparsity in the two-dimensional spatial domain. Additionally, a semi-definite programming (SDP) method with low computational complexity is proposed to solve the atomic minimization problem. Finally, both simulation and prototype are carried out to show estimation performance, and the proposed method outperforms the existing methods in the two-dimensional DOA estimation with low complexity in the scenario with practical RIS.
\end{abstract}
 
\begin{IEEEkeywords}
DOA estimation, practical RIS, atomic norm, mutual coupling, sparse reconstruction.
\end{IEEEkeywords}

\section{Introduction}
\IEEEPARstart{R}{econfigurable} intelligent surface (RIS) or intelligent reflecting surface (IRS) has been an attractive technology to improve the efficiency of spectrum and energy with a low-cost feature, and can control reflected signals and provide a configurable  wireless propagation environment~\cite{kang_irs_aided_2022,peng_channel_2022,a_l_swindlehurst_channel_2022}. Usually, the RIS is composed of multiple elements, and each element can control the amplitude or phase of the reflected signals. The RIS is a passive plane with a varactor diode and metamaterial structure without high-cost radio-frequency (RF)~\cite{tang_mimo_2020,wu_towards_2020}. Recently, an active RIS is also proposed by adding a power supply to provide additional reflection gain~\cite{fu_active_2022}. The RIS can be used to improve wireless communication and sensing performance:
\begin{itemize}
	\item For wireless communication applications, an IRS-aided communication system is proposed in~\cite{zheng_simultaneous_2022} to achieve transmitting diversity and passive beamforming. Two-sided cooperative IRSs are introduced in a low-earth orbit (LEO) satellite communication~\cite{zheng_intelligent_2022} to improve the achievable rate. 
	\item For the applications of sensing targets, the RIS can provide an additional path from targets to a sensor~\cite{y_wang_reconfigurable_2022,j_he_beyond_2022}, especially in the non-line-of-sight (NLOS) scenario. With the improved coverage, the RIS-based radar system is developed in~\cite{a_aubry_ris_aided_2021} to sense the targets. In~\cite{z_wang_location_2022,y_jiang_reconfigurable_2022}, a RIS-aided localization for near and far filed is formulated, where the localization performance is improved significantly by the RIS. Additionally, a Swendsen-Wang sampling-based method is proposed in~\cite{a_parchekani_sensing_2022} to optimize the RIS scanning channels for the target localization. 
\end{itemize}
 
To sense the targets, direction-of-arrival (DOA) estimation is an important way to obtain the directions of targets in the scenario with multiple receiving antennas, where the information on phase differences among antennas is exploited~\cite{kulkarni_non_integer_2022,zhengRadarCommunicationCoexistence2019,chuSuperResolutionMmWaveChannel2019,zheng_subttd_2022}. The traditional DOA estimation methods can be categorized into non-parametric and parametric methods. For the non-parametric methods, the fast Fourier transformation (FFT) method, Welch's method~\cite{welch_use_1967}, periodogram method~\cite{auger_improving_1995}, etc. have been proposed. For the parametric methods, autoregressive moving average (ARMA) method, multiple signal classification (MUSIC) method~\cite{schenck_probability_2022}, maximum entropy spectral estimation method~\cite{abramovich_positive_definite_1998}, etc. have been proposed. Recently, some new methods have been studied for DOA estimation. Deep learning (DL)-based method as a data-driven way is shown in~\cite{ma_deep_2022}. A radial basis function neural network-based method is proposed in~\cite{zheng_neural_2022} for the DOA estimation and is trained by the location information data. In~\cite{su_deep_2022}, the deep unfolding networks are adopted for the DOA estimation for the nested array. Additionally, a gridless DL method based on a convolutional neural network is proposed in~\cite{wu_gridless_2022} using the Toeplitz covariance matrix structure. The sparse Bayesian learning-based DOA estimation algorithm~\cite{liu_robust_2022,zhang_doa_2022,z_chen_robust_2020} can achieve better estimation performance with relatively high computational complexity. Furthermore, in~\cite{su_underdetermined_2022}, an under-determined blind method is proposed in the scenario without prior knowledge of the array configuration. A DOA estimation method for the array with more targets than the array elements is proposed in~\cite{yadav_underdetermined_2022} by utilizing the structure and geometry of the difference coarray. In~\cite{zheng_structured_2022}, a tensor-based method is proposed for the coherent DOA estimation in the two-dimensional scenario by exploiting the tensor structure-property. By exploiting the spatial sparsity of DOA, a decoupled atomic norm minimization (DANM)-based algorithm is shown in~\cite{mao_joint_2022} to estimate the DOA and range jointly.

Using the RIS, the targets can be sensed efficiently with the DOA estimation. In~\cite{k_ardah_trice_2021}, a channel estimation method based on orthogonal matching pursuit (OMP) is proposed by exploiting the low-rank nature of millimeter-wave multiple-input and multiple-output (MIMO) systems~\cite{guo_double_2022}, where the RIS is used to provide an indirect link between the base station and the mobile station. In~\cite{p_chen_reconfigurable_2021}, for the DOA estimation problem in a non-uniform linear array (ULA), a transformation matrix is introduced in the atomic norm minimization-based estimation method by exploiting the signal sparsity in the spatial domain. A DL-based method for the DOA estimation is proposed in~\cite{wan_deep_2021} using RIS and a deep unfolding network. A practical localization system leveraging on RIS is proposed in~\cite{a_albanese_papir_2021} with the help of RIS to localize the user equipment. Moreover, in~\cite{chen_efficient_2022}, the RIS is lifted in an unmanned aerial vehicle (UAV), and a DOA estimation method is proposed by solving the problem of UAV perturbation by a customized semi-definite programming (SDP) step. The controlling weights in the RIS are optimized in~\cite{z_yang_low-cost_2022} to optimize the beamforming scheme under the sparse reconstruction constraint at the DOA estimation step. Additionally, ref.~\cite{x_wang_joint_2022} optimizes the RIS phase shifts under the constraint of DOA estimation performance. 

However, when the RIS is adopted to estimate the direction of the target, the estimation performance will be degraded seriously in the scenario with RIS model mismatch~\cite{basarReconfigurableIntelligentSurfaceBased2020}. Recently, some estimation and optimization methods have been proposed in the practical scenario where the imperfect RIS is adopted. For example, for the near-field localization problem, the impact of hardware impairments in the RIS is studied in~\cite{ozturkImpactHardwareImpairments2022}. A deep reinforcement learning method for optimizing the phase shift in the RIS with practical consideration is given in~\cite{hashemiDeepReinforcementLearning2022}. The RIS phase shift errors affect the secrecy performance of the wireless communication~\cite{salemImpactPhaseShiftError2022}. Additionally, A simple way to model the mutual coupling of reconfigurable surfaces is shown in~\cite{direnzoModelingMutualCoupling2022}. However, for the high-accuracy two-dimensional problem, the DOA estimation method needs to be better studied in the practical scenario.

In this paper, a two-dimensional DOA estimation problem is investigated in the scenario using the RIS, where a sensor with only one full-functional channel is adopted to receive the reflected signals. Additionally, with the practical hardware, both the mutual coupling effect and the reflection phase/amplitude errors are considered in the system model. Then, a novel DOA estimation method is proposed by combining the deep neural network (DNN) and the decoupling atomic norm minimization (DANM). To decrease the computational complexity of the SDP problem at the atomic norm minimization step, a novel SDP problem is formulated. Finally, a prototype RIS is adopted to show the DOA estimation performance in the practical scenario. The contributions of this paper are summarized as follows:
\begin{itemize}
	\item \textbf{The system model for practical RIS is formulated.} For the two-dimensional DOA estimation problem, a system model using the RIS is formulated, where only one full-functional channel is adopted in the receiving sensor. The description of the reflected signals considers both the mutual coupling effect among adjacent elements and the phase/amplitude errors.
	\item \textbf{A novel DOA estimation method is proposed for the practical RIS.} The proposed method, DNN-DANM, combines the DNN and DANM steps to reconstruct the reflected signal from the practical RIS. The DNN step takes the ideal reflection signals as references, and the DANM step exploits the signal sparsity in the spatial domain. Additionally, the existing methods to minimize the atomic norm have high complexity in the two-dimensional scenario, so we proposed an improved SDP method to decrease the computational complexity in the sparse reconstruction step.
	\item \textbf{The DOA estimation performance is carried out by a prototype RIS.} A prototype RIS is processed and welded to show the proposed method's DOA estimation performance. Then, the existing DOA estimation methods are compared with the proposed method in different scenarios.
\end{itemize}
The advantage of the proposed method in improving the DOA estimation performance is that the signal affected by the RIS impairments can be reconstructed with high precision by the DNN, and the DANM can estimate the two-dimensional DOA accurately with low computational complexity. Hence, in the low-cost system with two-dimensional RIS and one full-functional receiving channel, the proposed method can achieve better performance in the DOA estimation than the existing methods.

\textit{Notations:} The  matrices and column vectors are denoted by upper-case and lower-case boldface letters, respectively. $(\cdot)^\text{H}$ is the  matrix Hermitian, and the transpose is denoted as $(\cdot)^\text{T}$. $\mathcal{R}\{\cdot\}$ and $\mathcal{I}\{\cdot\}$ denote the real and imaginary part of a complex value, respectively. $\text{Tr}\{\cdot\}$ is the trace of a matrix.  $\|\cdot\|_1$ and $\|\cdot\|_2$ are  the $\ell_1$ and $\ell_2$ norms, respectively.  $\mathbb{C}$ is the set of complex numbers, and we also define $j\triangleq\sqrt{-1}$. $\operatorname{vec}\{\boldsymbol{A}\}$ is the vectorization of a matrix $\boldsymbol{A}$, and converts the matrix into a column vector.

\section{The RIS-Based System Model for the DOA Estimation }\label{system}
\begin{figure}
	\centering
	\includegraphics[width=3.3in]{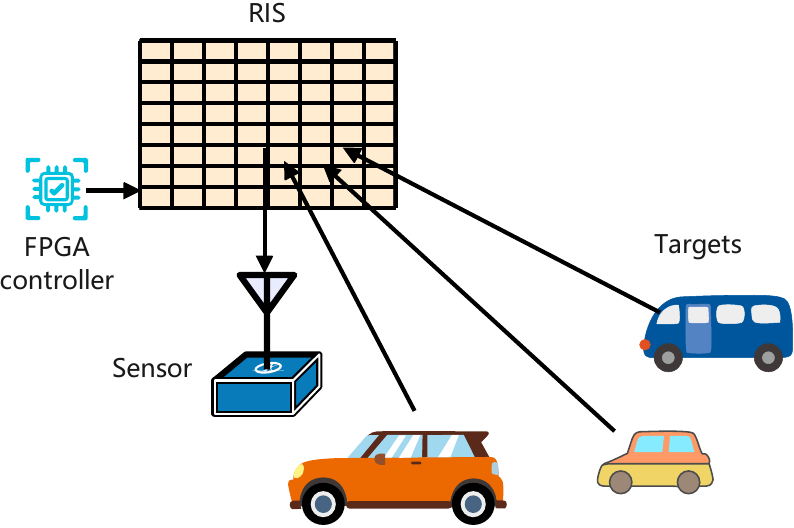}
	\caption{The RIS-based sensing model.}
	\label{fig1}
\end{figure} 

\begin{figure}
	\centering
	\includegraphics[width=3in]{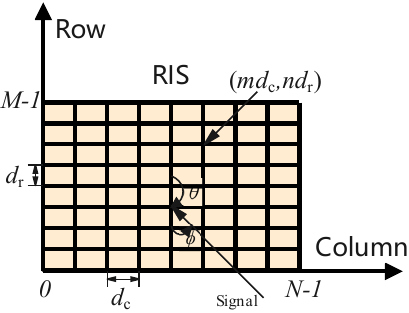}
	\caption{The geometrical topology of the RIS and received signal.}
	\label{fig2}
\end{figure}

In traditional systems,  multiple receiving channels are used to estimate the DOA from the phase information of the receiving signals among the channels. The systems using multiple channels are complex and high-cost, so they cannot be used widely and limit the sensing applications. This paper considers a RIS-based target sensing problem, as shown in Fig.~\ref{fig1}. There are multiple targets, and a RIS reflects the target signals to a sensor. With only one full-functional receiving channel in the sensor, we are trying to estimate DOA and obtain the target directions. Different from the traditional antenna array, The RIS has the following characteristics:
\begin{itemize}
	\item There is no RF channel and baseband processor in the RIS, so the RIS has a low-cost and cannot process the receiving signals.
	\item  The reflection coefficients in the RIS can be set as some  discrete values, so the signal control precision is relatively low.
	\item Due to the low cost of RIS and the complex installation environment, more non-ideal hardware will be used, resulting in signal distortion.
\end{itemize}
The proposed system model can expand the scope of sensing applications since both the RIS and the one-channel sensor are low-cost equipment.

The geometrical topology of the RIS and received signal is shown in Fig.~\ref{fig2}. We consider a two-dimensional RIS, where the RIS has $M$ rows and $N$ columns. The position of RIS element at the $m$-th ($m=0,1,\dots,M-1$) row and the $n$-th ($n=0,1,\dots, N-1$) column is $(md_{\text{r}}, nd_{\text{c}})$, where $d_{\text{c}}$ is the horizontal distance between the  adjacent RIS elements and $d_{\text{c}}$ is the corresponding vertical distance. We consider estimating the directions of $K$ far-field narrowband signals to locate the targets, where the azimuth and elevation of the $k$-th ($k=0,1,\dots,K-1$) signal are $\phi_k$ ($\phi_k\in[-\ang{90},\ang{90}]$) and $\theta_k$ ($\theta_k\in[\ang{0},\ang{180}]$), respectively. Taking the RIS element at the position ($0$, $0$) as a reference, the received signal by the RIS element at $(md_{\text{r}}, nd_{\text{c}})$ can be expressed as
\begin{align} 
r_{m,n}(t) = \sum_{k=0}^{K-1}e^{j\psi_{m,n}(\theta_k,\phi_k)}s_k(t),
\end{align}
where $s_{k}(t)$ is the $k$-th narrowband signal, the phase is defined as 
\begin{align}
\psi_{m,n}(\theta_k,\phi_k)\triangleq \frac{-2\pi}{\lambda}(nd_{\text{c}} \sin\theta_k\sin\phi_k +md_{\text{r}} \cos\theta_k),
\end{align}
and $\lambda$ denotes the wavelength.

To estimate the two-dimensional direction (azimuth and elevation), a FPGA is adopted in the RIS to control the amplitudes and phases of the reflections.  Then, the reflected signal at $(md_{\text{r}}, nd_{\text{c}})$ can be written as 
\begin{align}
z_{m,n}(t) = A_{m,n}(t)e^{j\zeta_{m,n}(t)}r_{m,n}(t),
\end{align}
where $A_{m,n}(t)e^{j\zeta_{m,n}(t)}$ is the reflection coefficient, $A_{m,n}(t)\in[0,1]$ is the amplitude, and $\zeta_{m,n}(t)\in[0,2\pi)$ is the phase. Since the power of the reflected signal is less than that of the incident signal, it is reasonable to assume that $A_{m,n}\leq 1$. Then, a sensor with only one receiving channel is placed at the front of the RIS with the azimuth being $\ang{0}$ and the elevation being $\ang{90}$. The sensor can receive the reflected signal, so the received signal at the sensor can be obtained as
\begin{align}\label{eq3}
    y(t) = \sum_{m=0}^{M-1}\sum_{n=0}^{N-1}z_{m,n}(t)+w(t),
\end{align}
where $w(t)$ is the additive white Gaussian noise. 

With the sampling interval $T$ and the narrowband assumption, the received signal in (\ref{eq3}) can be rewritten by a vector form as
\begin{align}\label{eq4}
    \boldsymbol{y}=\boldsymbol{GAs}+\boldsymbol{w},
\end{align}
where we define
\begin{align}
\boldsymbol{y}&\triangleq\left[y(0), y(T),\dots, y((P-1)T)\right]^{\text{T}}\in\mathbb{C}^{P\times 1},\\
\boldsymbol{w}&\triangleq\left[w(0), w(T),\dots, w((P-1)T)\right]^{\text{T}}\in\mathbb{C}^{P\times 1},
\end{align} 
and $P$ is the number of samples. Additionally, the steering matrix $\boldsymbol{A}\in \mathbb{C}^{MN\times K}$ in (\ref{eq4}) is defined as
\begin{align}
\boldsymbol{A}&\triangleq \left[\boldsymbol{a}(\theta_0,\phi_0),\boldsymbol{a}(\theta_1,\phi_1),\dots,\boldsymbol{a}(\theta_{K-1},\phi_{K-1})\right],
\end{align}
and the steering vector $\boldsymbol{a}(\theta,\phi)\in\mathbb{C}^{MN\times 1}$ is 
\begin{align}\label{eq_st}
	\boldsymbol{a}(\theta,\phi)& \triangleq \left[e^{j\psi_{0,0}(\theta,\phi)},\dots, e^{j\psi_{M-1,N-1}(\theta,\phi)}\right]^{\text{T}}.
\end{align}
The signals from $K$ targets is defined as
\begin{align}
\boldsymbol{s}&\triangleq \left[s_{0},s_{1},\dots, s_{K-1}\right]^{\text{T}}\in\mathbb{C}^{K\times 1}.
\end{align}
The reflection coefficients are collected into a reflection matrix, and can be expressed as
\begin{align}
\boldsymbol{G}& \triangleq \left[\boldsymbol{g}(0),\boldsymbol{g}(T),\dots,\boldsymbol{g}((P-1)T)\right]^{\text{T}},
\end{align}
where the $p$-th column of the reflection matrix $\boldsymbol{G}\in\mathbb{C}^{P\times MN}$ is
\begin{align} 
\boldsymbol{g}(pT)&\triangleq\left[g_{0,0}(pT),\dots, g_{M-1,M-1}(pT)\right]^{\text{T}}\in\mathbb{C}^{MN\times 1},
\end{align}
and $g_{m,n}(pT)\triangleq A_{m,n}(pT)e^{j\zeta(pT)}$. 

The system model in (\ref{eq4}) is for the ideal scenario. However, 
for practical consideration, the RIS cannot perfectly control the reflection process, and there are mismatches in describing the reflection coefficients. Typically, considering a 1-bit RIS, the reflection coefficient can be written as
\begin{align}
    g_{m,n}(pT)=\begin{cases}
    1,&b_{m,n}(pT)=0\\
    -1,&b_{m,n}(pT)=1
    \end{cases},
\end{align}
where $b_{m,n}(pT)$ is the FPGA control signal during the $p$-th interval. With the reflection mismatch, we can take the reflection coefficient at $b_{m,n}(pT)=0$ as reference, and we have
\begin{align}
    g_{m,n}(pT)=\begin{cases}
    1,&b_{m,n}(pT)=0\\
    -B_{m,n}e^{j\beta_{m,n}},&b_{m,n}(pT)=1
    \end{cases}, 
\end{align}
where we use  $B_{m,n}\in[0,1]$ and $\beta_{m,n}\in[0,2\pi)$ to describe the reflection mismatch with the control signal being $1$. 

Moreover, the RIS differs from the traditional antenna array, and the distance between the adjacent elements is less than half of the wavelength. The ideal array assumption in (\ref{eq4}) that the elements receive the signals independently is impractical, and the mutual coupling effect among elements must be considered. Hence, a mutual coupling matrix $\boldsymbol{C}\in\mathbb{C}^{MN\times MN}$ can be used to describe the mutual coupling effect among the RIS elements. The entry at the $(mN+n)$-th row and the $(m'N+n')$-th column of $\boldsymbol{C}$ gives the mutual coupling coefficient between the RIS element at ($md_{\text{r}}$, $nd_{\text{c}}$) and that at ($m'd_{\text{r}}$, $n'd_{\text{c}}$). Additionally, the diagonal entries of $\boldsymbol{C}$ are ones.

Therefore, considering both the mismatch of the reflection coefficient and the mutual coupling effect in the practical $1$-bit RIS, the received signal in the sensor can be finally rewritten as
\begin{align}\label{eq16}
\tilde{\boldsymbol{y}}=(\boldsymbol{B}\odot\boldsymbol{G})\boldsymbol{CAs}+\boldsymbol{w}
\end{align}
where $\odot$ is the Hadamard product, the entry of reflection matrix $\boldsymbol{G}$ is $1$ or $-1$, and the matrix $\boldsymbol{B}$ describes the reflection mismatch. The entry of $\boldsymbol{B}$ is $1$ when that of $\boldsymbol{G}$ at the same position is $1$, and is $-B_{m,n}e^{j\beta_{m,n}}$  when that of $\boldsymbol{G}$  is $-1$. The matrix $\boldsymbol{C}$ describes the mutual coupling effect among the RIS elements.  In this paper, we try to estimate the azimuth and elevation directions of targets from the received signal $\boldsymbol{y}$ using the RIS and the sensor in the scenario with the unknown signal $\boldsymbol{s}$, the reflection mismatch $\boldsymbol{B}$ and the mutual coupling effect $\boldsymbol{C}$.

\section{The DOA Estimation Method With Practical RIS}

\begin{figure}
	\centering
	\includegraphics[width=2.5in]{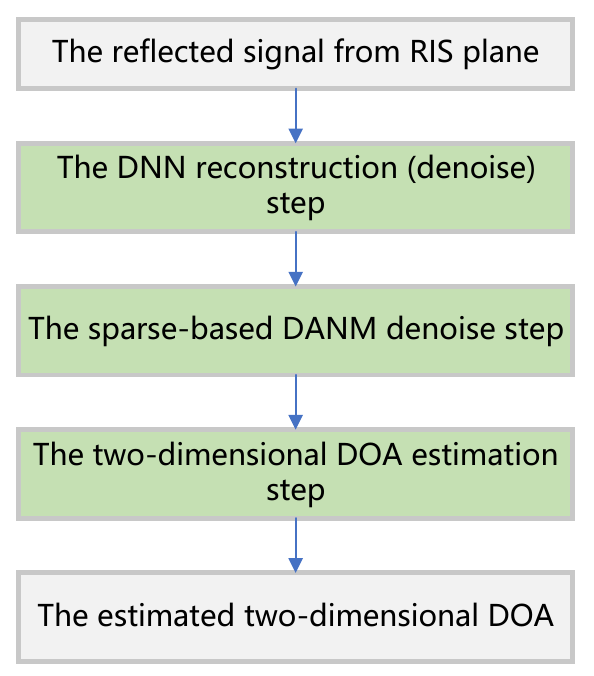}
	\caption{The steps for the proposed DNN-DANM method to estimate the DOA.}
	\label{step}
\end{figure} 

In order to accurately estimate the two-dimensional DOA in the practical RIS-based system, the key step is to reduce the impact of the non-ideal RIS. This paper proposes a DNN-DANM method combing a deep neural network (DNN) and a decoupling atomic norm minimization (DANM) to reconstruct the received signal and estimate the two-dimensional DOA. As shown in Fig.~\ref{step}, there are $3$ steps in the DNN-DANM method, i.e., the DNN-based reconstruction step, the DANM-based denoising step, and the two-dimensional DOA estimation step. We will show the details of these steps.

\subsection{The DNN-Based Reconstruction Step}
\begin{figure}
	\centering
	\includegraphics[width=3.7in]{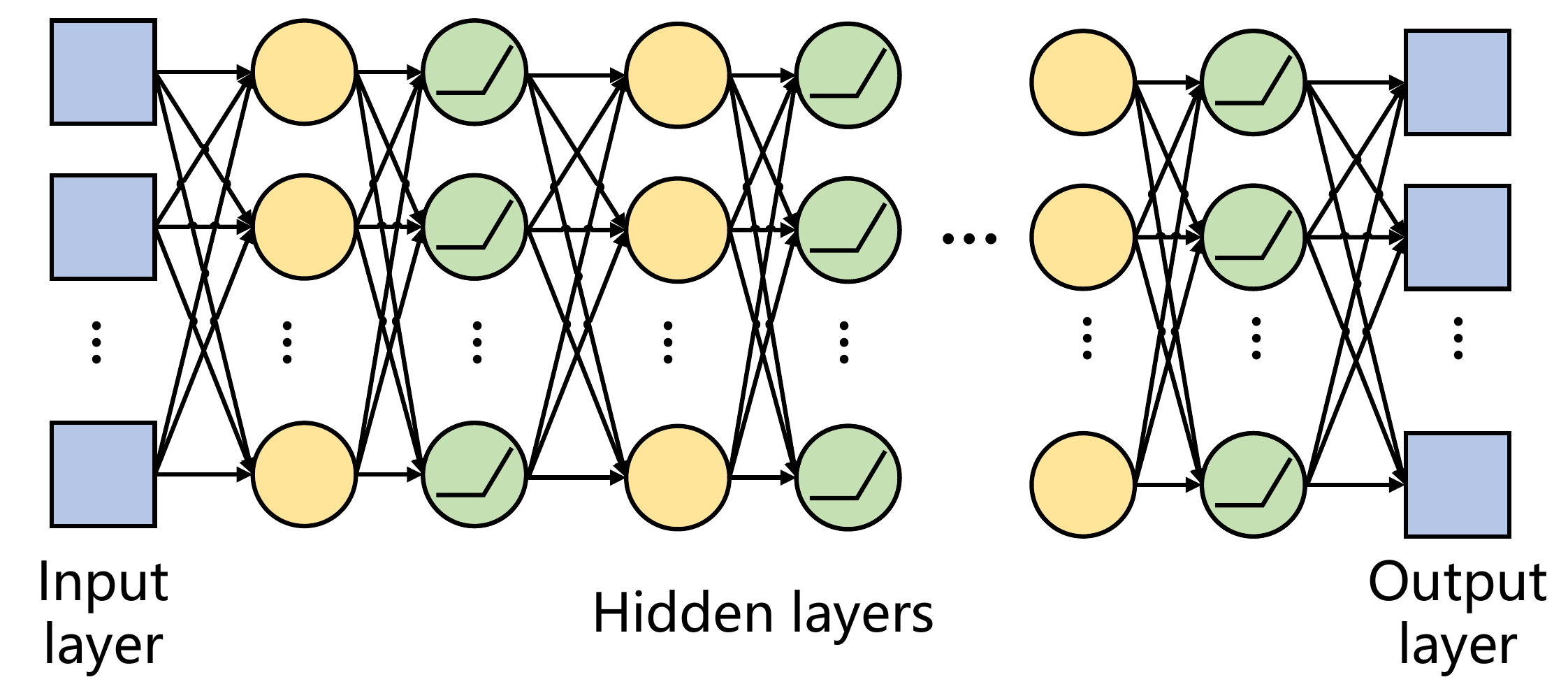}
	\caption{The DNN structure for the signal reconstruction with system model mismatch.}
	\label{net}
\end{figure} 

When the RIS-based system is adopted to estimate the directions of targets, the reflection mismatch and the mutual coupling effect degrade the estimation performance significantly. Hence, in the first step of the proposed DNN-DANM method, we try to reconstruct the signal from the received one with the system model mismatch. 

As shown in Fig.~\ref{net}, a fully connected network is adopted to reconstruct the signal. The received signal in (\ref{eq16}) can be rewritten as
\begin{align} 
\tilde{\boldsymbol{y}}_{\text{RI}}\triangleq [\mathcal{R}\{\tilde{\boldsymbol{y}}\}^{\text{T}}, \mathcal{I}\{\tilde{\boldsymbol{y}}\}^{\text{T}}]^{\text{T}}\in\mathbb{R}^{2P\times 1},
\end{align}
which is used as the DNN input. The DNN output denoted as $\boldsymbol{z}_{\text{RI}}\triangleq [\mathcal{R}\{\boldsymbol{z}\}^{\text{T}},\mathcal{I}\{\boldsymbol{z}\}^{\text{T}}]^{\text{T}}$ is the reconstructed signal and has the same size as the input. In the hidden layers, a ReLU function is adopted in the activation layer and is defined as
\begin{align}
f_{\text{ReLU}}(x)=\begin{cases}
x,&\quad x\geq 0\\
0,&\quad x<0
\end{cases}.
\end{align}

To train the DNN, the following loss function is used
\begin{align}
f_{\text{Loss}}(\boldsymbol{z}_{\text{RI}})=\frac{1}{2P}\|\boldsymbol{z}_{\text{RI}}-\boldsymbol{y}_{\text{RI}}\|^2_2,
\end{align}
where we define $\boldsymbol{y}_{\text{RI}}\triangleq [\mathcal{R}\{\boldsymbol{y}\}^{\text{T}},\mathcal{I}\{\boldsymbol{y}\}^{\text{T}}]^{\text{T}}$ as the reference signal with the perfect system model. To obtain the perfect signals, we can use the system model in (\ref{eq4}), where the geometrical topology of the RIS and the reflection matrix are used to generate the ideal signals. The reflection mismatch and the mutual coupling effect are not considered in the perfect system model. By training the network, the output signal is the reconstructed signal, and can approach the one with the perfect system model.  

\subsection{The DANM-Based Denoising Step}
Since the RIS is a type of low-cost equipment, the number of RIS elements is much larger than that of targets, i.e., $MN\gg K$. After the network, the following reconstructed signal 
\begin{align}
\boldsymbol{z}\triangleq \mathcal{R}\{\boldsymbol{z}\}+j\mathcal{I}\{\boldsymbol{z}\}
\end{align}
can be obtained. Then, an atomic norm minimization method can be further applied to remove noise from the signal by exploiting the signal sparsity in the spatial domain.

We can formulate the following optimization problem
\begin{align}\label{eq9}
    \min_{\boldsymbol{x}}\frac{1}{2}\|\boldsymbol{z}-\boldsymbol{Gx}\|^2_2+\alpha\|\boldsymbol{x}\|_{\mathcal{A},0},
\end{align}
where the parameter $\alpha$ is adopted to control the balance between the reconstruction error and the sparsity. $\|\cdot\|_{\mathcal{A},0}$ is defined as~\cite{chenNewAtomicNorm2020}
\begin{equation} 
	\begin{split}
\|\boldsymbol{x}\|_{\mathcal{A},0}\triangleq \inf_{Q}\Bigg\{\boldsymbol{x}&=\sum_{q=0}^{Q-1}c_qe^{j\psi_q}\boldsymbol{a}(\theta_q,\phi_q),\\
& c_q\geq 0, \psi_q\in[0,2\pi)\Bigg\}.
	\end{split}
\end{equation}
However, the optimization problem (\ref{eq9}) is non-deterministic polynomial-time (NP)-hard, and cannot be solved efficient. Hence, a convex relaxation of $\|\boldsymbol{x}\|_{\mathcal{A},0}$ is introduced and called the atomic norm of $\boldsymbol{x}$, which is defined as
\begin{equation}\label{eq11}
	\begin{split}
\|\boldsymbol{x}\|_{\mathcal{A}}\triangleq \inf\Bigg\{\sum_{q=0}^{Q-1}c_q\Big|\boldsymbol{x}&=\sum_{q=0}^{Q-1}c_qe^{j\psi_q}\boldsymbol{a}(\theta_q,\phi_q),\\
& c_q\geq 0, \psi_q\in[0,2\pi)\Bigg\}.
	\end{split}
\end{equation}
Then, the optimization problem (\ref{eq9}) can be relaxed as 
\begin{align} 
    \min_{\boldsymbol{x}}\frac{1}{2}\|\boldsymbol{z}-\boldsymbol{Gx}\|^2_2+\alpha\|\boldsymbol{x}\|_{\mathcal{A}}.\label{eq12}
\end{align}
  
To solve the atomic norm-based optimization problem, we have the following proposition:
\begin{theorem}\label{th1}
To obtain the atomic norm defined in (\ref{eq11}), the following optimization problem is formulated 
\begin{align}
\min_{\boldsymbol{T},t} & \frac{1}{2}\left(\operatorname{Tr}\{\boldsymbol{T}\}+t\right)\\
\text{s.t. }& \begin{bmatrix}
\boldsymbol{T} & \boldsymbol{x}\\
\boldsymbol{x}^{\text{H}} & t
\end{bmatrix}\succeq 0\notag\\
&  \boldsymbol{T}= \begin{bmatrix}
\boldsymbol{T}_0 & \boldsymbol{T}_1,&\dots&\boldsymbol{T}_{N-1}\\
\boldsymbol{T}^{\text{H}}_1 & \boldsymbol{T}_0&\dots&\boldsymbol{T}_{N-2}\\
\vdots & \vdots &\ddots&\vdots\\
\boldsymbol{T}^{\text{H}}_{N-1} & \boldsymbol{T}^{\text{H}}_{N-2}  &\dots &  \boldsymbol{T}_{0}
\end{bmatrix}\in\mathbb{C}^{MN\times MN}\notag\\
&\boldsymbol{T}_n=
\begin{bmatrix}
T_{n,0} & T_{n,1}   &\dots &T_{n,M-1}\\
T_{n,-1} & T_{n,0}  &\dots &T_{n,M-2}\\ 
\vdots &   \vdots &\ddots&\vdots\\
T_{n,1-M} & T_{n,2-M}  &\dots &T_{n,0}
\end{bmatrix}\in\mathbb{C}^{M\times M}\notag\\
& \boldsymbol{T}\text{ is a block Hermitian Toeplitz matrix}\notag\\
& \boldsymbol{T}_n\text{ is a Toeplitz matrix}\notag
\end{align}
If the optimal $\boldsymbol{T}$ can be written as $
    \boldsymbol{T} = \boldsymbol{U}\boldsymbol{D}\boldsymbol{U}^{\text{H}}$,
where $\boldsymbol{D} $ is a diagonal matrix with then diagonal entries being real and positive, and $\boldsymbol{U}=\begin{bmatrix}
\boldsymbol{a}(\theta_0,\phi_0),\boldsymbol{a}(\theta_1,\phi_1),\dots \end{bmatrix}$, the atomic norm can be obtained as the objective value of the optimization problem $ \|\boldsymbol{x}\|_{\mathcal{A}} =  \frac{1}{2}\left(\operatorname{Tr}\{\boldsymbol{T}\}+t\right)$.
\end{theorem}

The proof of  Proposition~\ref{th1} is given in Appendix~\ref{ap1}.With Proposition~\ref{th1}, the signal after removing the noise can be obtained by the atomic norm minimization (ANM) problem (\ref{eq12}). We can find that the size of the positive semi-definite matrix $\begin{bmatrix}
\boldsymbol{T} &\boldsymbol{x}\\
\boldsymbol{x}^{\text{H}}  &t
\end{bmatrix}$ is $(MN+1)\times (MN+1)$, so the computational complexity of solving the ANM problem is $\mathcal{O}((MN+1)^{3.5})$. The computational complexity is too high and is not efficient in the scenario using a RIS with large size. 

To solve the problem of the high computational complexity, a DANM-based method is proposed to remove the noise, and we have the following proposition~\cite{zhang_efficient_2019}
\begin{theorem}\label{thm2}
For the two-dimensional DOA estimation, the ANM problem in (\ref{eq12}) can be obtained from the following SDP problem
\begin{align}\label{eq32}
\min_{\boldsymbol{T}_{\text{x}},\boldsymbol{T}_{\text{y}},\boldsymbol{X}}\quad& \frac{1}{2}\left(\operatorname{Tr}\{\boldsymbol{T}_{\text{x}}\}+\operatorname{Tr}\{\boldsymbol{T}_{\text{y}}\}\right)\\
\text{s.t.}\quad &\begin{bmatrix}
\boldsymbol{T}_{\text{x}} &\boldsymbol{X}\\
\boldsymbol{X}^{\text{H}} &\boldsymbol{T}_{\text{y}}
\end{bmatrix}\succeq 0\notag\\
& \|\boldsymbol{z}-\boldsymbol{G}\operatorname{vec}\{\boldsymbol{X}\}\|_2\leq \sqrt{P_{\text{n}}}\notag\\
&\boldsymbol{T}_{\text{x}}\in\mathbb{C}^{M\times M} \text{ is a Hermitian Toeplitz matrix}\notag\\
&\boldsymbol{T}_{\text{y}}\in\mathbb{C}^{N\times N} \text{ is a Hermitian Toeplitz matrix}\notag\\
&\boldsymbol{X}\in\mathbb{C}^{M\times N},\notag
\end{align}
where $P_{\text{n}}$ is the noise power.
\end{theorem}
The proof of Proposition~\ref{thm2} is given as follows:
\begin{proof} 
From Theorem~4.2 in~\cite{zhang_efficient_2019}, we can find that the two-dimensional atomic norm of a matrix $\boldsymbol{X}\in\mathbb{C}^{M\times N}$ can be obtained as
\begin{equation}
\begin{split}
\|\boldsymbol{X}\|_{\mathcal{A}}=&\min_{\boldsymbol{T}_{\text{x}},\boldsymbol{T}_{\text{y}}}\left\{\frac{1}{2\sqrt{MN}}\left(\operatorname{Tr}\{\boldsymbol{T}_{\text{x}}\}+\operatorname{Tr}\{\boldsymbol{T}_{\text{y}}\}\right)\right\}\\
&\text{s.t. }\begin{bmatrix}
	\boldsymbol{T}_{\text{x}} & \boldsymbol{X}\\
	\boldsymbol{X}^{\text{H}} & \boldsymbol{T}_{y}
\end{bmatrix}\succeq 0\\
&\quad \boldsymbol{T}_{\text{x}}\text{ and } \boldsymbol{T}_{\text{y}}\text{ are the Toeplitz matrices,}
\end{split}
\end{equation}
where the atomic norm of $\boldsymbol{X}$ is defined as
\begin{equation}
\begin{split}
\|\boldsymbol{X}\|_{\mathcal{A}}\triangleq \inf \Bigg\{\left.\sum_{q=0}^{Q-1}c_qe^{j\psi_q}\right|\boldsymbol{X}&=\sum_{q=0}^{Q-1}c_q\bar{\boldsymbol{a}}(f_1)\bar{\bar{\boldsymbol{a}}}^{\text{H}}(f_2),\\
& c_q\geq 0,\psi_q\in[0,2\pi)\Bigg\},
\end{split}	
\end{equation}
and the steering vectors are defined as
\begin{equation}
\begin{split}
\bar{\boldsymbol{a}}(f_1)\triangleq [1, e^{-j\frac{2\pi}{\lambda}d_{\text{r}}f_1},\dots, e^{-j\frac{2\pi}{\lambda}(M-1)d_{\text{r}}f_1}]^{\text{T}},\\
\bar{\bar{\boldsymbol{a}}}(f_2)\triangleq [1, e^{-j\frac{2\pi}{\lambda}d_{\text{c}}f_2},\dots, e^{-j\frac{2\pi}{\lambda}(N-1)d_{\text{c}}f_2}]^{\text{T}}.
\end{split}	
\end{equation}

From the definition of steering vector in (\ref{eq_st}) with $\phi\in[-\pi/2,\pi/2]$ and $\theta\in[0,\pi]$, let
$f_1\triangleq \sin\theta\sin\phi \in[-1,1]$ and $f_2\triangleq  \cos\theta\in[-1,1]$, and we can find that
\begin{equation}
\begin{split}
	\boldsymbol{a}(\theta,\phi) &= \left[e^{j\psi_{0,0}(\theta,\phi)},\dots, e^{j\psi_{M-1,N-1}(\theta,\phi)}\right]^{\text{T}}\\ 
	& =  \left[\dots,e^{\frac{-j2\pi}{\lambda}(nd_{\text{c}}f_1 +md_{\text{r}}f_2)},\dots\right]^{\text{T}}\\
	& = \operatorname{vec}\left\{\bar{\boldsymbol{a}}(f_1)\bar{\bar{\boldsymbol{a}}}^{\text{H}}(f_2)\right\}.
\end{split}
\end{equation}
Therefore, we can obtain
\begin{align}
\|\boldsymbol{X}\|_{\mathcal{A}}=\|\operatorname{vec}\{\boldsymbol{X}\}\|_{\mathcal{A}},
\end{align}
and the atomic norm minimization of $\operatorname{vec}\{\boldsymbol{X}\}$ is equal to that of $\boldsymbol{X}$. Then, with the additive white Gaussian noise, the optimization problem in (\ref{eq32}) can be obtained.
\end{proof}

We can find that the size of the matrix $\begin{bmatrix}
	\boldsymbol{T}_{\text{x}} &\boldsymbol{X}\\
	\boldsymbol{X}^{\text{H}} &\boldsymbol{T}_{\text{y}}
\end{bmatrix}$ in Proposition~\ref{thm2} is $(M+N)\times(M+N)$, so the computational complexity to solve the SDP problem (\ref{eq32}) is $\mathcal{O}((M+N)^{3.5})$, which is much lower than that in Proposition~\ref{th1} in the scenario using RIS with large number of elements.

\subsection{The Two-Dimensional DOA Estimation Step}
After the DNN-based reconstruction and the ANM-based removing noise steps, the influence of the system mismatch can be effectively reduced. Finally, we can obtain the  two-dimensional DOA.

From the SDP problem (\ref{eq32}), we can obtain $\boldsymbol{T}_{\text{x}}$ and $\boldsymbol{T}_{\text{y}}$. The first column of $\boldsymbol{T}_{\text{x}}$ is denoted as $\boldsymbol{u}_{\text{x}}$, and the $m$-th entry is denoted as $u_{\text{x},m}$. Then, we can formulate a Toeplitz matrix as
\begin{align}
\boldsymbol{H}_{\text{x}} \triangleq \begin{bmatrix}
u_{\text{x}, K-1} & u_{\text{x}, K-2} & \dots & u_{\text{x}, 0}\\
u_{\text{x}, K} & u_{\text{x}, K-1} & \dots & u_{\text{x}, 1}\\
\vdots & \vdots & \ddots & \vdots\\
u_{\text{x}, M-2} & u_{\text{x}, M-3} & \dots & u_{\text{x}, K-1}
\end{bmatrix}.
\end{align}
A vector $\boldsymbol{b}$ can be obtained as
\begin{align}
\boldsymbol{b}=\boldsymbol{H}^{\dagger}_{\text{x}}\boldsymbol{u}_{\text{x}, K:M-1}.
\end{align}
Hence, a polynomial with the coefficients being $\bar{\boldsymbol{b}}\triangleq [1, \boldsymbol{b}^{\text{T}}]^{\text{T}}$ can be formulated, and the roots can be obtained as
\begin{align}
\boldsymbol{v}=\texttt{roots}(\bar{\boldsymbol{b}}),
\end{align}
where $\texttt{roots}(\bar{\boldsymbol{b}})$ is a function to get the roots of the polynomial represented by a vector $\bar{\boldsymbol{b}}$.  
Finally, the azimuth can be estimated as 
\begin{align}
\hat{\boldsymbol{\phi}} = \arccos\left(-\frac{\angle{\boldsymbol{v}}}{2\pi d_{\text{r}}}\right).
\end{align}

Similarly, from the matrix $\boldsymbol{T}_{\text{y}}$, the elevation can be estimated as $\hat{\boldsymbol{\theta}}$. After this step, the DOA of targets can be estimated accurately.

\section{Simulation Results}

\begin{figure}
	\centering
	\includegraphics[width=2.5in]{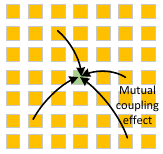}
	\caption{The system model for mutual coupling effect.}
	\label{mc}
\end{figure} 

In this section, simulation results are carried out in a personal computer (PC) with Intel(R) Core(TM) i7 processor and 16~GByte memory. The code for the proposed method is available online \url{https://github.com/chenpengseu/DNN-DANM.git}. 

\begin{figure}
	\centering
	\includegraphics[width=2.5in]{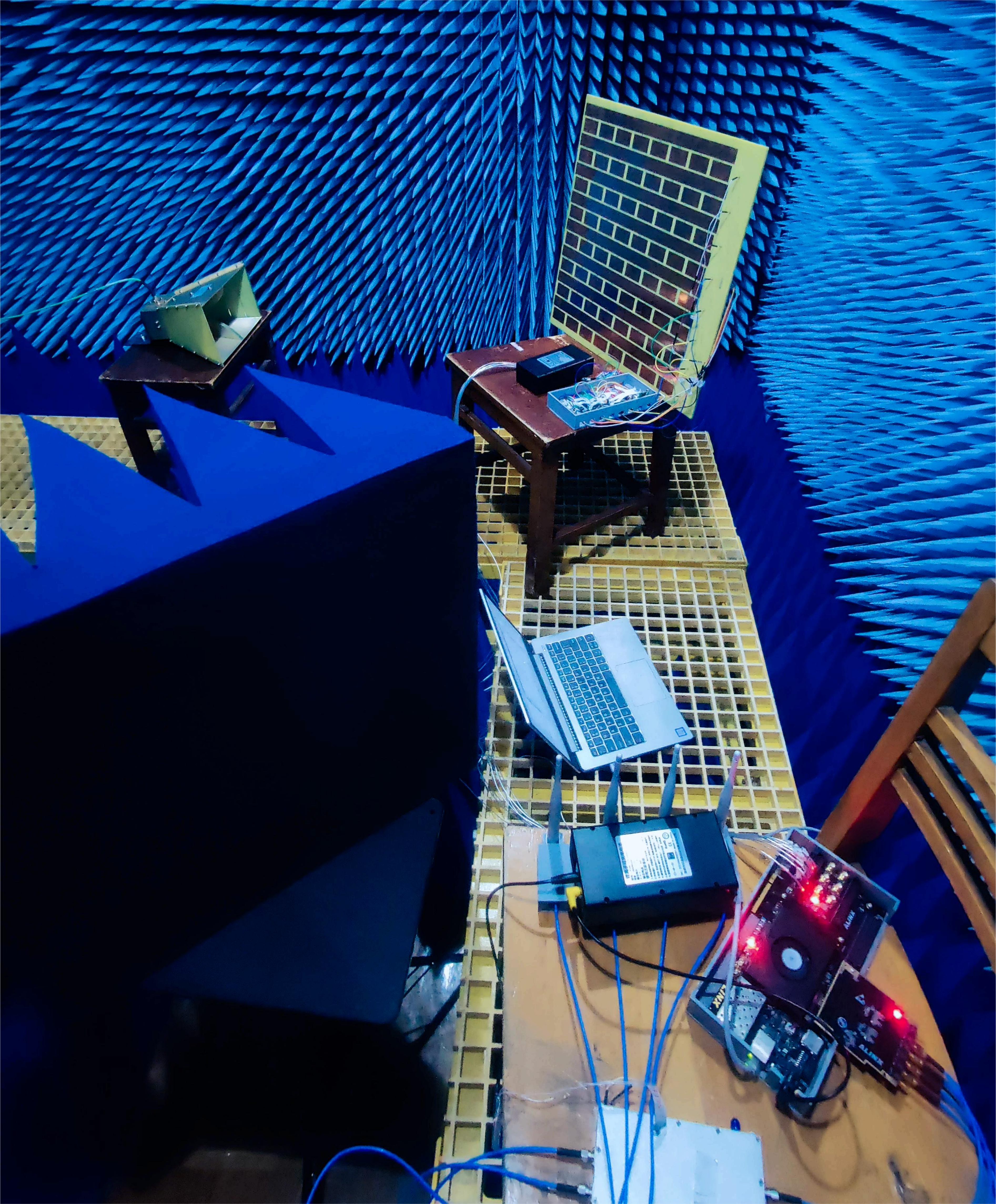}
	\caption{A practical RIS for the DOA estimation.}
	\label{img}
\end{figure}

Simulation parameters are given in Table~\ref{table1} to show the proposed method's performance in different scenarios. The two-dimensional DOAs of $2$ targets are estimated using a $16\times 16$ RIS, where the distance between adjacent RIS elements is $0.4\lambda$ and less than half of the wavelength. Hence, the mutual coupling effect must be considered, and the system model to describe the mutual coupling effect is shown in Fig.~\ref{mc}. The three adjacent elements affect the considering element, and the mutual coupling between the element at the $m$-th row and $n$-th column and that at the $m'$-th row and $n'$-th column can be written as $C_{mN+n,m'N+n'}$, whose amplitude and phase have the following uniform distribution
\begin{equation}
\begin{split}
|C_{mN+n,m'N+n'}| & \sim U_{[0.1, 0.4]},\\
\angle{C} &\sim U_{[0,2\pi)}.
\end{split}
\end{equation}
The mutual coupling matrix $\boldsymbol{C}$ is more general than existing methods using a symmetric Toeplitz matrix since the practical RIS is not perfectly symmetric, and the mutual coupling among the elements is not identical, especially for the edge RIS elements. For the reflection mismatch matrix $\boldsymbol{B}$, the amplitude and phase of the entries have the following uniform distribution
\begin{equation}
\begin{split}
B_{m,n}&\sim U_{[0.5,1.5]},\\
\beta_{m,n}&\sim U_{[-\pi/6,\pi/6]}.
\end{split}
\end{equation}
The parameters to describe the reflection mismatch and the mutual coupling effect is set according to a practical RIS, as shown in Fig.~\ref{img}. For the number of samples $P$, we set it as $128$. This parameter is also important for the DOA estimation, and we choose it according to the RIS size and the number of targets~\cite{aaa}. 

\begin{table}% [!t] 
	\renewcommand{\arraystretch}{1.3}
	\caption{Simulation parameters}
	\label{table1}
	\centering
	\begin{tabular}{cc}
		\hline
		\textbf{Parameter} & \textbf{Value}  \\
		\hline 
		The distance between the adjacent RIS elements & $d_{\text{c}}=d_{\text{r}}=0.4\lambda$\\
		The number of RIS rows & $M=16$\\
		The number of RIS columns & $N=16$\\
		The number of samples & $P=128$\\
		The number of targets & $K=2$\\
		The azimuth of target & $\phi_k\in[\ang{-30},\ang{30}]$ \\
		The elevation of target & $\theta_k\in[\ang{20},\ang{80}]$\\
		\hline
	\end{tabular}
\end{table}

\begin{figure}
	\centering
	\includegraphics[width=3in]{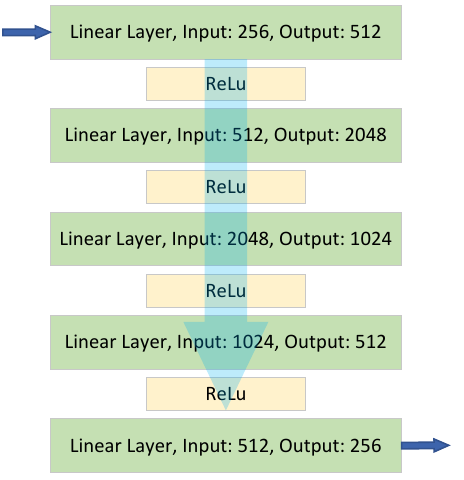}
	\caption{The DNN structure for the signal reconstruction.}
	\label{net-str}
\end{figure} 

\begin{figure}
	\centering
	\includegraphics[width=3.5in]{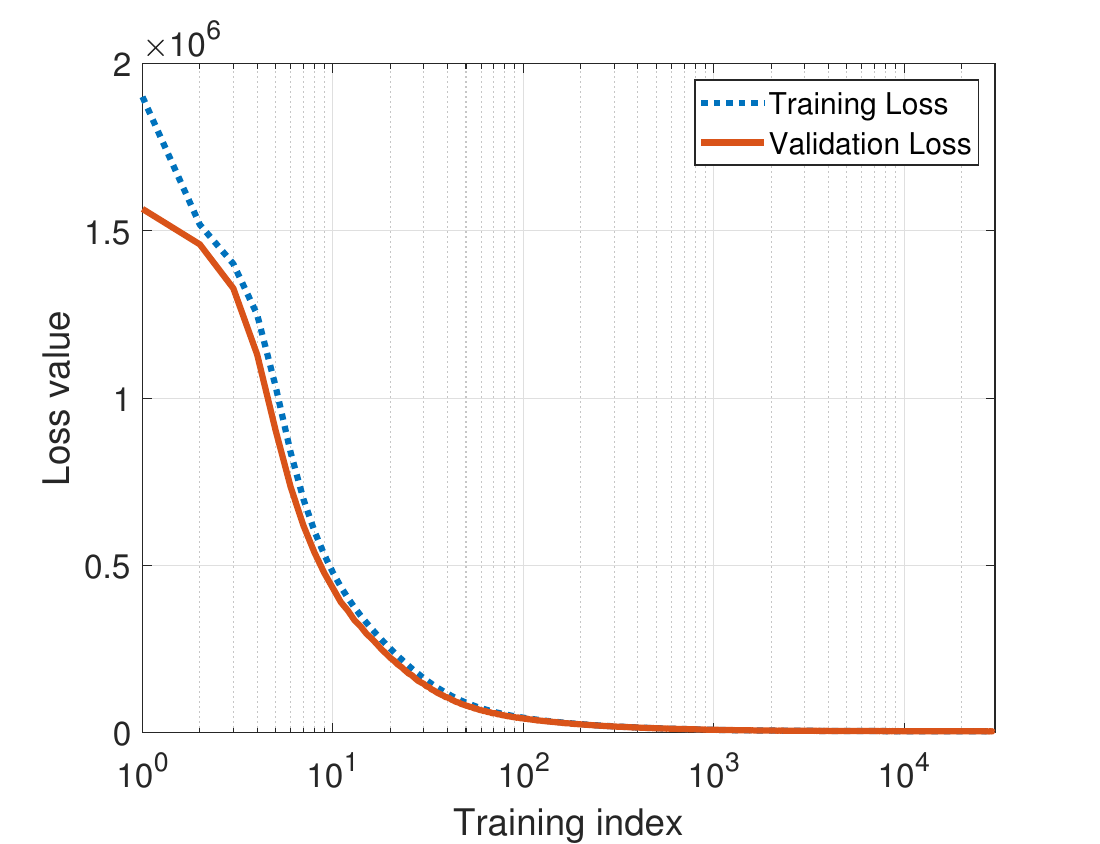}
	\caption{The values of the loss function in the training step.}
	\label{loss}
\end{figure} 

The DNN structure for the signal reconstruction is shown in Fig.~\ref{net-str}, and the input layer size is $2P$. After $5$ linear layers, the output size is also $2P$, which is the same as the input size. The back-propagation is used to update the network weights during the training step.

First, the values of the loss function to train the DNN and reconstruction the received signal is shown in Fig.~\ref{loss}, where the training data set is generated by changing the received signals in the scenarios with the SNR from $20$~dB to $50$~dB and choosing the values of both the mutual coupling effect and the reflection mismatch randomly. The Adam algorithm updates the network with the learning rate of $10^{-4}$, and the batch size is $64$. As shown in this figure, the loss function approaches zero when the number of training epochs is greater than $10^3$.

\begin{figure}
	\centering
	\includegraphics[width=3.5in]{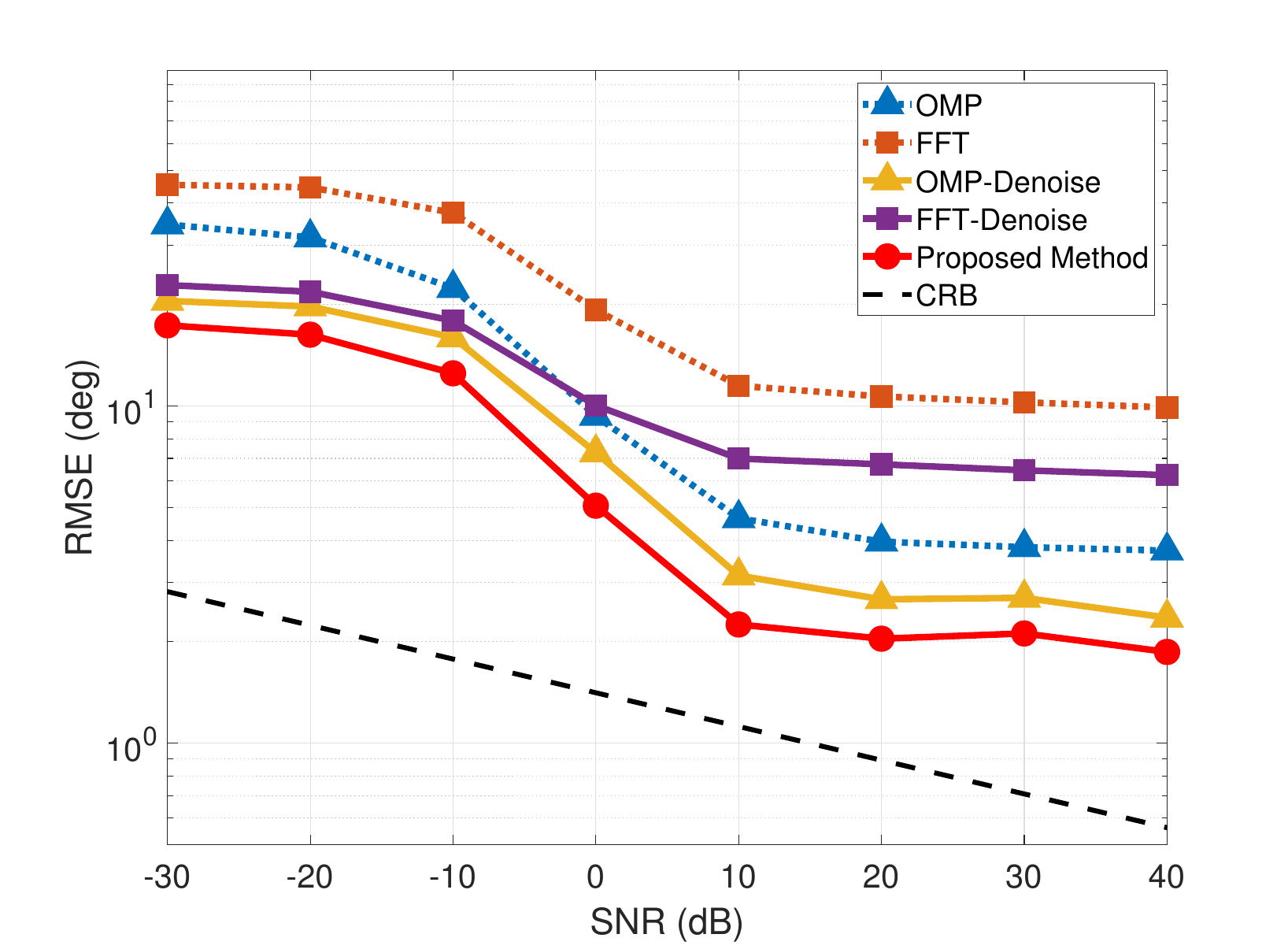}
	\caption{The DOA estimation performance with different SNRs.}
	\label{snr}
\end{figure} 

Then, after training the DNN, the proposed DNN-DANM method is adopted to show the two-dimensional DOA estimation performance in the scenario with different SNRs. As shown in Fig.~\ref{snr}, the proposed method is compared with the orthogonal matching pursuit (OMP)~\cite{linSingleSensorEstimate2021}, fast Fourier transform (FFT), OMP-denoise and FFT-denoise methods. The OMP-denoise and FFT-denoise methods use the output of the DNN reconstruction to estimate the target DOA. The estimation performance is measured by the root mean square error (RMSE), which is defined as
\begin{align}
e_{\text{MSE}}\triangleq \sqrt{\frac{1}{2N_{\text{mc}K}}\left(\left\|\hat{\boldsymbol{\theta}}-\boldsymbol{\theta}\right\|^2_2+\left\|\hat{\boldsymbol{\phi}}-\boldsymbol{\phi}\right\|^2_2\right)},
\end{align}
where we define the azimuth vector as $\boldsymbol{\phi}\triangleq [\phi_0,\dots,\phi_{K-1}]^{\text{T}}$ and the elevation vector as $\boldsymbol{\theta}\triangleq [\theta_0,\dots,\theta_{K-1}]^{\text{T}}$. $\hat{\boldsymbol{\theta}}$ and $\hat{\boldsymbol{\phi}}$ are the estimated $\boldsymbol{\theta}$ and $\boldsymbol{\phi}$, respectively. $N_{\text{mc}}$ is the number of Monte Carlo trails, and is set as $10^3$ in the our simulation. The Cram\'{e}r-Rao bound (CRB) is obtained from the results in~\cite{chen_efficient_2022}. As shown in this figure, the DNN reconstruction step can improve the estimation performance. For the FFT method, the RMSE is about $\ang{10}$ at $\text{SNR}=30$~dB, and the RMSE can be reduced to $\ang{7}$ with the DNN reconstruction. For the OMP method, the DOA estimation performance is improved by about $\ang{2}$. Additionally, the proposed DNN-DANM can improve the DOA estimation performance and is better than the OMP-denoise method. The proposed method can approach the CRB in the scenario with the SNR being $10$~dB. Since the reflection mismatch and the mutual coupling effect are considered, a performance gap exists between the proposed method's RMSE and CRB. Hence, the estimation performance cannot be further improved by setting the SNR greater than $20$~dB.

\begin{figure}
	\centering
	\includegraphics[width=3.5in]{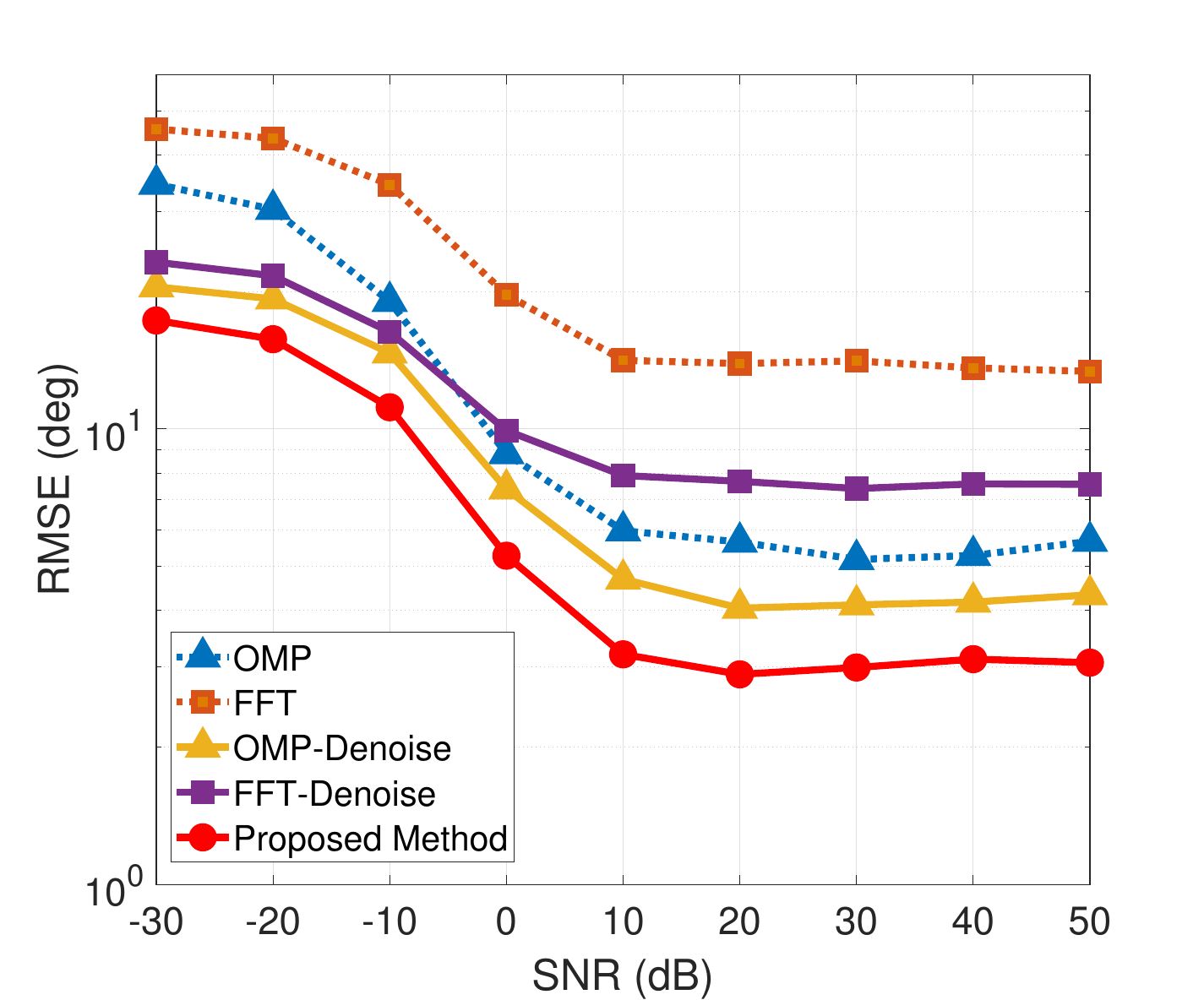}
	\caption{The DOA estimation performance with the amplitude error of the reflection coefficient being $B_{m,n}\sim U_{[0.5,3]}$.}
	\label{snr-amp}
\end{figure} 

Additionally, when the reflection mismatch is worse, and we change the distribution to $B_{m,n} \sim U_{[0.5,3]}$, the corresponding performance of the DOA estimation is shown in Fig.~\ref{snr-amp}. Compared with Fig.~\ref{snr}, the DOA estimation performance has deteriorated, and the RMSE of the proposed method is about $\ang{3}$ with $\text{SNR}=20$~dB. However, the proposed method outperforms FFT and OMP methods even in the scenario with the DNN reconstruction step. 
 
\begin{figure}
	\centering
	\includegraphics[width=3.5in]{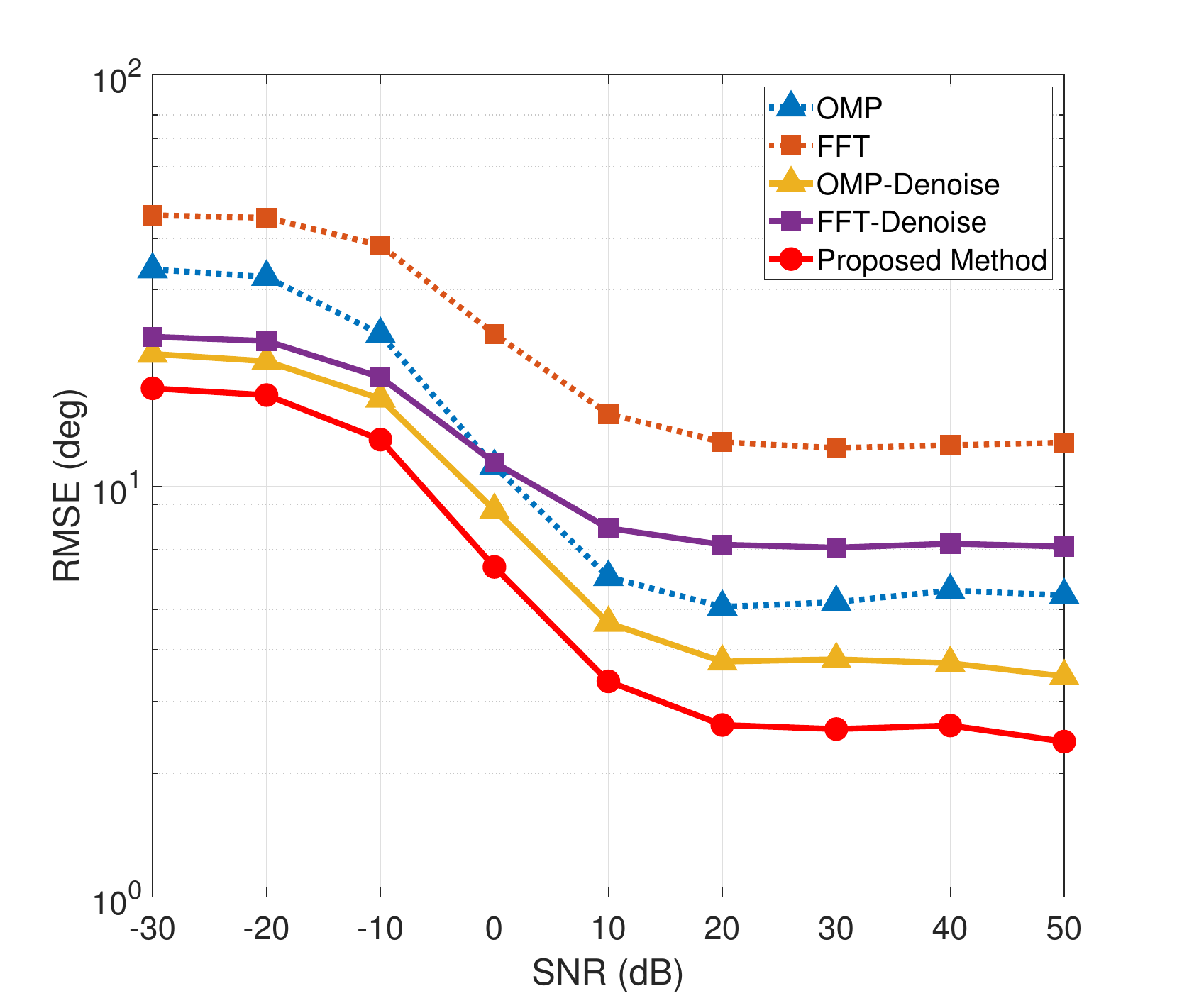}
	\caption{The DOA estimation performance with the phase error of the reflection coefficient being $\beta_{m,n}\sim U_{[-\pi/3,\pi/3]}$.}
	\label{snr-phase}
\end{figure} 

When we change the phase distribution of the reflection mismatch from $\beta_{m,n}\sim U_{[-\pi/6,\pi/6]}$ to $\beta_{m,n}\sim U_{[-\pi/3,\pi/3]}$, the RMSE of the DOA estimation is shown in Fig.~\ref{snr-phase}. The proposed DNN-DANM can perform better than the OMP, FFT, OMP-denoise, and FFT-denoise methods. Therefore, for the RIS with reflection mismatch in the system model, the DNN-DANM can outperform the traditional methods, especially in the scenario with large mismatch coefficients.

\begin{figure}
	\centering
	\includegraphics[width=3.5in]{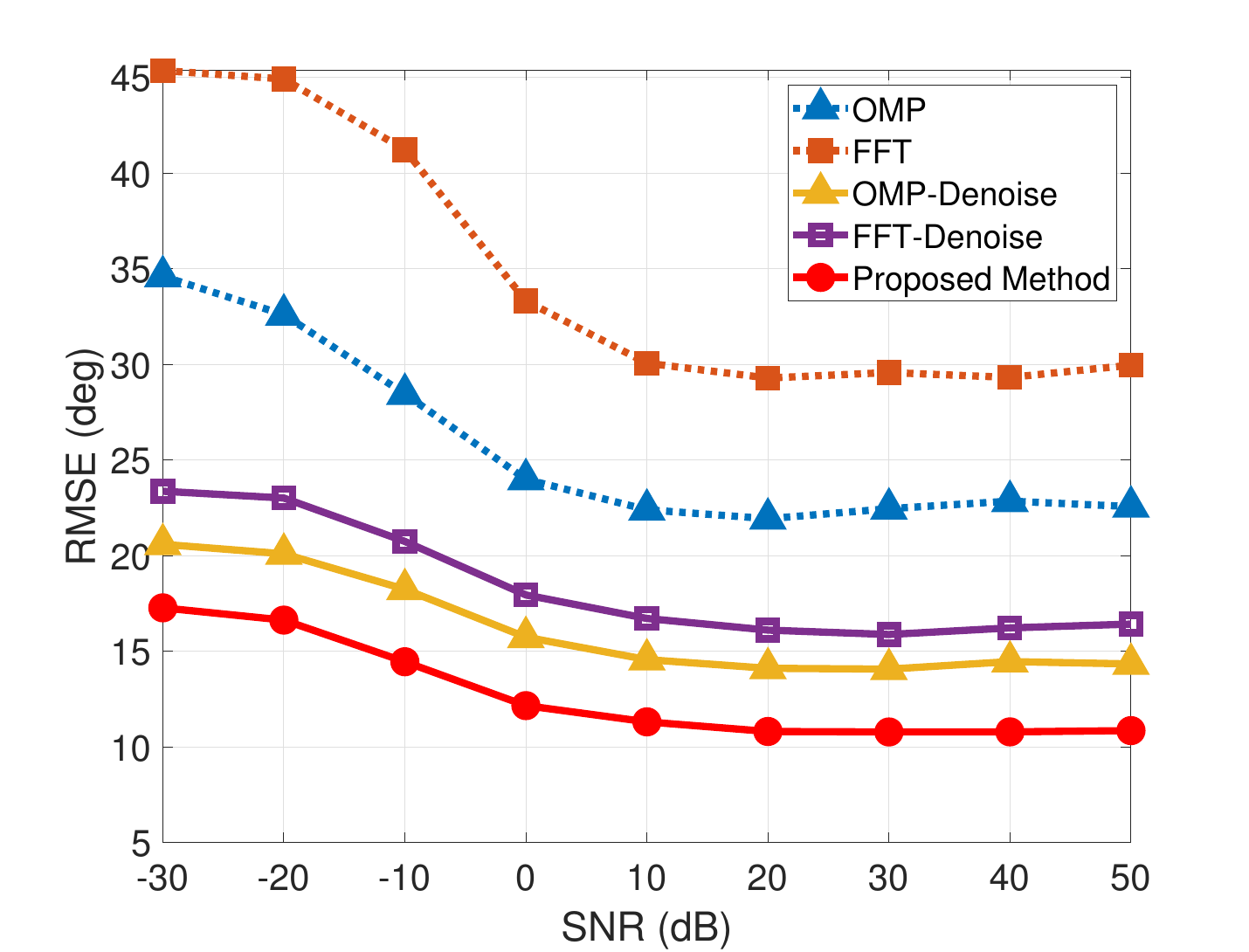}
	\caption{The DOA estimation performance with the mutual coupling coefficient being $|C_{m,n}|\sim U_{[0.1,0.8]}$.}
	\label{snr-mc}
\end{figure}

\begin{table}% [!t] 
	\renewcommand{\arraystretch}{1.3}
	\caption{Computing times}
	\label{table2}
	\centering
	\begin{tabular}{cc}
		\hline
		\textbf{Method} & \textbf{Time (s)}  \\
		\hline 
	OMP & $0.085$\\
	FFT & $0.035$\\
	OMP-Denoise & $0.199$\\
	FFT-Denoise & $0.172$\\
	ANM-Denoise & $171.703$\\
	The Proposed Method& $3.608$ \\ 
		\hline
	\end{tabular}
\end{table}

Finally, we show the DOA estimation performance in the scenario with a larger mutual coupling effect, and the estimation performance is shown in Fig.~\ref{snr-mc}, where the distribution of the mutual coupling effect is $|C_{m,n}|\sim U_{[0.1,0.8]}$. We can find that the mutual coupling effect degrades the DOA estimation significantly, but better estimation performance is achieved by the proposed DNN-DANM method. Therefore, from the simulation results, we can find that the proposed method is efficient in the DOA estimation with the RIS, especially in the scenario with the mismatch of the system model. To compare the computational complexity, the computing times of different algorithms are given in  Table~\ref{table2}. We can find that OMP-based and FFT-based methods have low computational complexity, but the estimation performance is limited. The ANM-denoise method in Proposition~\ref{th1} can achieve better estimation performance, but the computing time is $171.703$~s and is too complex. The proposed method can achieve a balance between the computational complexity and the estimation performance. The computing time of the proposed method is $3.608$~s, much lower than the ANM-denoise method.

\section{Conclusions} 
The two-dimensional DOA estimation problem has been investigated in the scenario with only one full-functional receiving channel, and the imperfect hardware with the mutual coupling effect and the reflection amplitude/phase has been considered. We have combined the DNN and the DANM to propose the novel DNN-DANM method to estimate the DOA accurately. Moreover, the low-complex atomic norm minimization method has been proposed for the two-dimensional DOA estimation. The DOA estimation performance of the proposed method outperforms the existing methods using the prototype RIS. Future work will focus on the DOA estimation for the wideband signals using the RIS.

\bibliographystyle{IEEEtran}
\bibliography{IEEEabrv.bib,ref.bib} 

\appendices
\section{Proof of Proposition~\ref{th1}} \label{ap1}
The proof of Proposition~\ref{th1} is given as follows:
\begin{proof}
	When we formulate $\boldsymbol{x}=\sum _q c_q e^{j\psi_q}\boldsymbol{a}(\theta_q,\phi_q)$, we can obtain
	\begin{align}
		& \sum_q c_q\begin{bmatrix}
			\boldsymbol{a}(\theta_q,\phi_q)\\
			e^{-j\psi_q}
		\end{bmatrix}\begin{bmatrix}
			\boldsymbol{a}(\theta_q,\phi_q)\\
			e^{-j\psi_q}
		\end{bmatrix}^{\text{H}}\\
		& =\sum_q c_q\begin{bmatrix}
			\boldsymbol{a}(\theta_q,\phi_q)\boldsymbol{a}^{\text{H}}(\theta_q,\phi_q) & \boldsymbol{a}(\theta_q,\phi_q)e^{j\psi_q}\\
			e^{-j\psi_q}\boldsymbol{a}^{\text{H}}(\theta_q,\phi_q) & 1
		\end{bmatrix}\notag\\
		& = \begin{bmatrix}\sum_q c_q
			\boldsymbol{a}(\theta_q,\phi_q)\boldsymbol{a}^{\text{H}}(\theta_q,\phi_q) & \boldsymbol{x}\\
			\boldsymbol{x}^{\text{H}} & \sum_q c_q
		\end{bmatrix}\succeq 0.
	\end{align}
	Since we can choose $\boldsymbol{T}=\sum_q c_q
	\boldsymbol{a}(\theta_q,\phi_q)\boldsymbol{a}^{\text{H}}(\theta_q,\phi_q)$ and $t=\sum_q c_q$, with the definition of atomic norm, we can obtain
	\begin{align}\label{eq15}
		\|\boldsymbol{x}\|_{\mathcal{A}} & = \sum_q c_q\\
		& = \frac{1}{2}\left(\operatorname{Tr}\{\sum_q c_q
		\boldsymbol{a}(\theta_q,\phi_q)\boldsymbol{a}^{\text{H}}(\theta_q,\phi_q)\}+\sum_q c_q\right)\notag\\
		&\geq \frac{1}{2}\left(\min_{\boldsymbol{T},t} \operatorname{Tr}\{\boldsymbol{T}\}+t\right).\notag
	\end{align}
	
	% We have
	% \begin{align}
		% & \boldsymbol{a}(\theta,\phi)\boldsymbol{a}^{\text{H}}(\theta,\phi) = e^{j\psi_{m,n}(\theta,\phi)}e^{-j\psi_{m',n'}(\theta,\phi)}\\
		% & = e^{j((n'-n)d_{\text{c}} \sin\theta\sin\phi +(m'-m)d_{\text{r}} \cos\theta)} 
		% \end{align}
	Formulating a block Hermitian Toeplitz matrix $\boldsymbol{T}\in\mathbb{C}^{MN\times MN}$, we have
	\begin{align}
		\boldsymbol{T}= \begin{bmatrix}
			\boldsymbol{T}_0 & \boldsymbol{T}_1 &\boldsymbol{T}_2&\dots&\boldsymbol{T}_{N-1}\\
			\boldsymbol{T}^{\text{H}}_1 & \boldsymbol{T}_0&\boldsymbol{T}_1&\dots&\boldsymbol{T}_{N-2}\\
			\boldsymbol{T}^{\text{H}}_2 & \boldsymbol{T}^{\text{H}}_1&\boldsymbol{T}_0&\dots&\boldsymbol{T}_{N-3}\\
			\vdots & \vdots & \vdots &\ddots&\vdots\\
			\boldsymbol{T}^{\text{H}}_{N-1} & \boldsymbol{T}^{\text{H}}_{N-2} & \boldsymbol{T}^{\text{H}}_{N-3} &\dots &  \boldsymbol{T}_{0}
		\end{bmatrix},
	\end{align}
	where the sub-matrix $\boldsymbol{T}_n\in\mathbb{C}^{M\times M}$ is a Toeplitz matrix and is defined as
	\begin{align}
		\boldsymbol{T}_n=
		\begin{bmatrix}
			T_{n,0} & T_{n,1} & T_{n,2} &\dots &T_{n,M-1}\\
			T_{n,-1} & T_{n,0} & T_{n,1} &\dots &T_{n,M-2}\\
			T_{n,-2} & T_{n,-1} & T_{n,0} &\dots &T_{n,M-3}\\
			\vdots & \vdots & \vdots &\ddots&\vdots\\
			T_{n,1-M} & T_{n,2-M} & T_{n,3-M} &\dots &T_{n,0}
		\end{bmatrix}.  
	\end{align}
	
	For the matrix $\boldsymbol{T}$ and parameter $t$, if we have
	\begin{align}
		\begin{bmatrix}
			\boldsymbol{T} & \boldsymbol{x}\\
			\boldsymbol{x}^{\text{H}} & t
		\end{bmatrix}\succeq 0,
	\end{align}
	we can obtain that $\boldsymbol{T}\succeq 0$ and $\boldsymbol{T}\succeq t^{-1}\boldsymbol{x}\boldsymbol{x}^{\text{H}}$ with the Schur complement condition. 
	
	If the matrix $\boldsymbol{T}$ can be written as
	\begin{align}
		\boldsymbol{T} = \boldsymbol{U}\boldsymbol{D}\boldsymbol{U}^{\text{H}},
	\end{align}
	where $\boldsymbol{D}\in\mathbb{C}^{Q'\times Q'}$ is a diagonal matrix with then diagonal entries being real and positive, and $\boldsymbol{U}=\begin{bmatrix}
		\boldsymbol{a}(\theta_0,\phi_0),\boldsymbol{a}(\theta_1,\phi_1),\dots, \boldsymbol{a}(\theta_{Q'-1},\phi_{Q'-1})\end{bmatrix}$, $\boldsymbol{x}$ can be rewritten as
	$\boldsymbol{x}=\boldsymbol{U}\boldsymbol{\zeta}$ with a vector $\boldsymbol{\zeta}$. Hence, we can obtain
	\begin{align}\label{eq20}
		\frac{1}{2}\operatorname{Tr}\{\boldsymbol{T}\}+\frac{1}{2}t&=\frac{1}{2}\operatorname{Tr}\{\boldsymbol{D}\}+\frac{1}{2}t\\
		&=\frac{1}{2}\boldsymbol{\zeta}'^{\text{H}}\boldsymbol{UDU}^{\text{H}}\boldsymbol{\zeta}'+\frac{1}{2}t\notag\\
		&\geq \frac{1}{2t}\boldsymbol{\zeta}'^{\text{H}}\boldsymbol{xx}^{\text{H}}\boldsymbol{\zeta}'+\frac{1}{2}t  \notag\\
		&= \frac{1}{2t}\boldsymbol{\zeta}'^{\text{H}}\boldsymbol{U}\boldsymbol{\zeta}(\boldsymbol{U}\boldsymbol{\zeta})^{\text{H}}\boldsymbol{\zeta}'+\frac{1}{2}t  \notag\\
		&= \frac{1}{2t}\operatorname{sign}^{\text{H}}(\boldsymbol{\zeta}) \boldsymbol{\zeta}
		\boldsymbol{\zeta}^{\text{H}}
		\operatorname{sign}(\boldsymbol{\zeta})+\frac{1}{2}t  \notag\\
		&= \frac{1}{2t} \|\boldsymbol{\zeta}\|_1^2+\frac{1}{2}t  \notag\\
		&\geq \sqrt{\|\boldsymbol{\zeta}\|_1^2}=\|\boldsymbol{\zeta}\|_1\geq \|\boldsymbol{x}\|_{\mathcal{A}},  \notag
	\end{align}
	where $\boldsymbol{U}^{\text{H}}\boldsymbol{\zeta}'=\operatorname{sign}(\boldsymbol{\zeta})$ and $\operatorname{sign}(\boldsymbol{\zeta})$ is the sign vector of $\boldsymbol{\zeta}$.
	
	Therefore, with (\ref{eq15}) and (\ref{eq20}), we can find that
	\begin{align}
		\|\boldsymbol{x}\|_{\mathcal{A}}=\frac{1}{2}\min_{\boldsymbol{T},t}\left( \operatorname{Tr}\{\boldsymbol{T}\}+t\right).
	\end{align} 
\end{proof}

\begin{IEEEbiography}[{\includegraphics[width=1in,height=1.25in,clip,keepaspectratio]{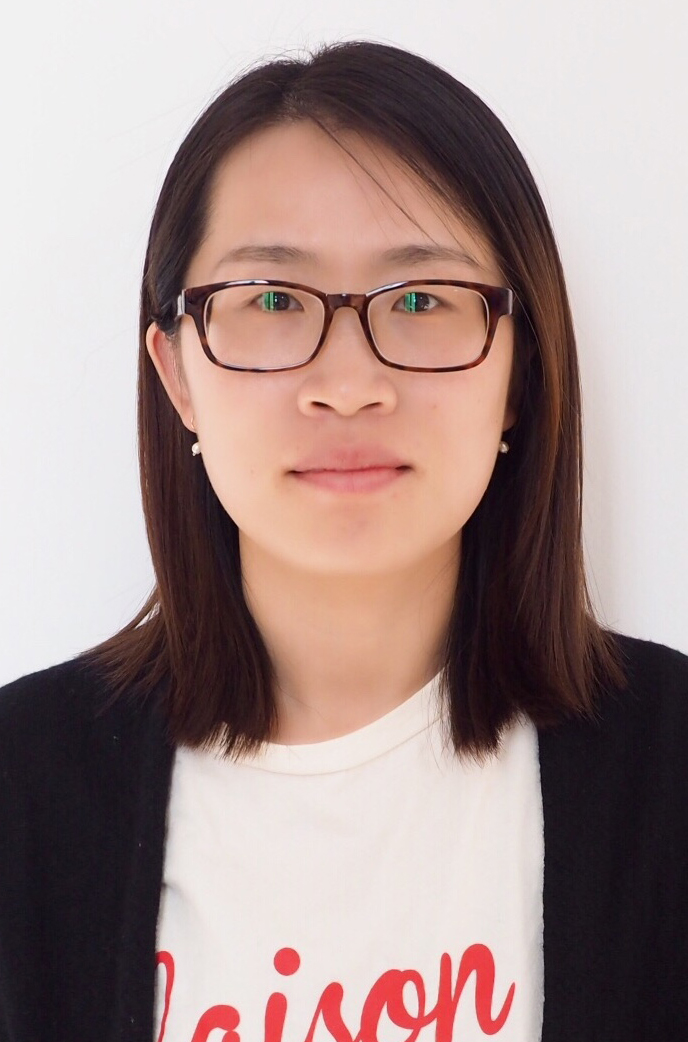}}]{Zhimin Chen (M'17)} received the Ph.D. degree in information and communication engineering from the School of Information Science and Engineering, Southeast University, Nanjing, China in 2015. Since 2015, she is currently an associate professor at Shanghai Dianji University, Shanghai, China. From 2021,  she is also a Visiting Scholar in  the Department of Electronic and Information Engineering, The Hong Kong Polytechnic University, Hong Kong.  Her research interests include array signal processing, vehicle communications and millimeter-wave communications. 
\end{IEEEbiography}

\begin{IEEEbiography}[{\includegraphics[width=1in,height=1.25in,clip,keepaspectratio]{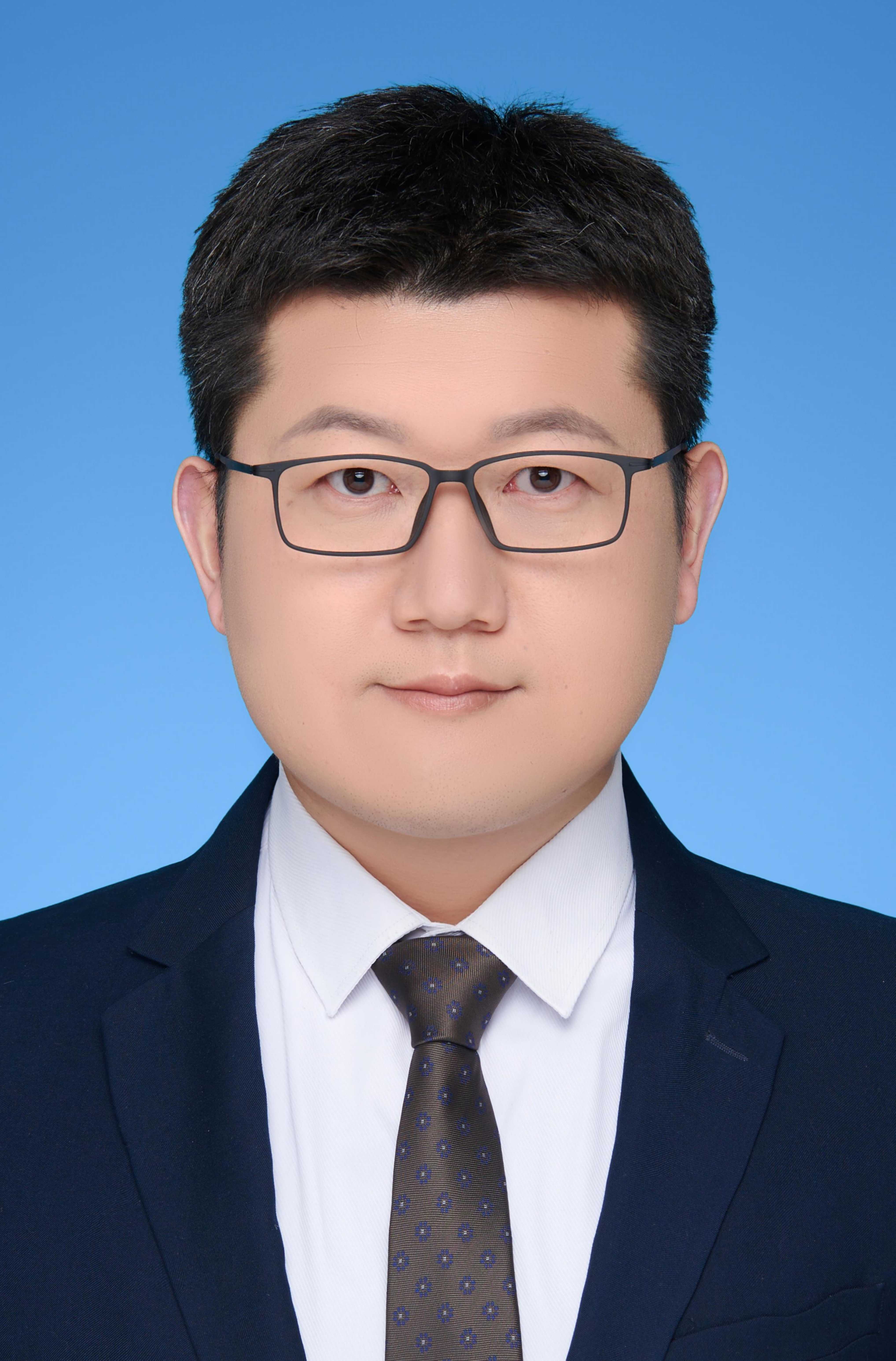}}]{Peng Chen (S'15-M'17-SM'22)} received the B.E. and Ph.D. degrees from the School of Information Science and Engineering, Southeast University, Nanjing, China, in 2011 and 2017 respectively. From March 2015 to April 2016, he was a Visiting Scholar with the Department of Electrical Engineering, Columbia University, New York, NY, USA. He is currently an Associate Professor with the State Key Laboratory of Millimeter Waves, Southeast University. His research interests include target localization, super-resolution reconstruction, and array signal processing. He is a Jiangsu Province Outstanding Young Scientist. He has served as an IEEE ICCC Session Chair, and won the Best Presentation Award in 2022 (IEEE ICCC). He was invited as a keynote speaker at the IEEE ICET in 2022. He was recognized as an exemplary reviewer for IEEE WCL in 2021, and won the Best Paper Award at IEEE ICCCCEE in 2017.
\end{IEEEbiography}

\begin{IEEEbiography}[{\includegraphics[width=1in,height=1.25in,clip,keepaspectratio]{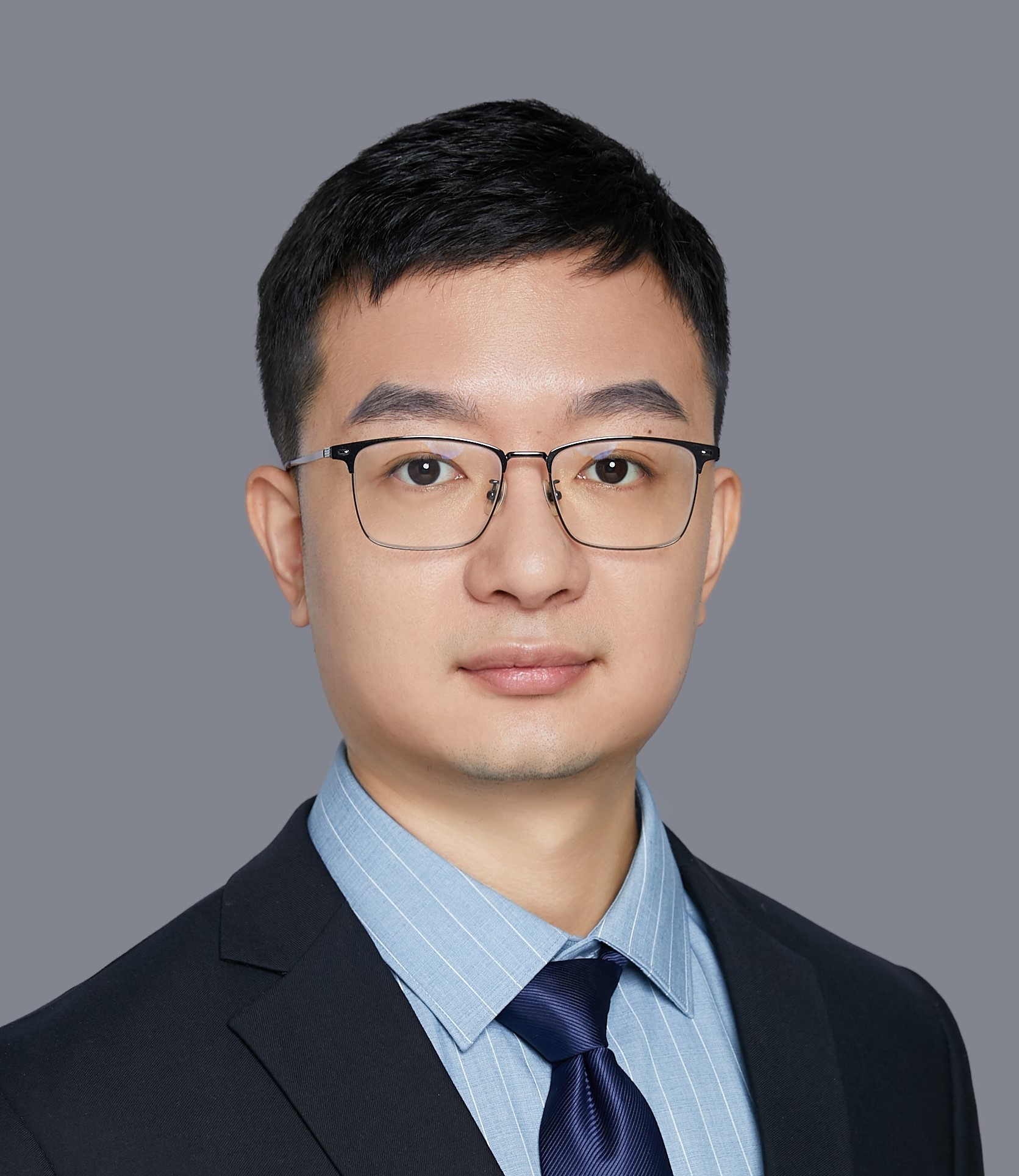}}]{Le Zheng (Senior Member, IEEE)} received the B.Eng. degree from Northwestern Polytechnical University (NWPU), Xi’an, China, in 2009 and Ph.D degree from Beijing Institute of Technology (BIT), Beijing, China in 2015, respectively. He has previously held academic positions in the Electrical Engineering Department of Columbia University, New York, U.S., first as a Visiting Researcher from 2013 to 2014 and then as a Postdoc Research Fellow from 2015 to 2017. From 2018 to 2022, he worked at Aptiv (formerly Delphi), Los Angeles, as a Principal Radar Systems Engineer, leading projects on the next-generation automotive radar products. Since 2022, he has been a Full Professor with the School of Information and Electronics, BIT. His research interests lie in the general areas of radar, statistical signal processing, wireless communication, and high-performance hardware, and in particular in the area of automotive radar and integrated sensing and communications (ISAC). 
\end{IEEEbiography}

\begin{IEEEbiography}[{\includegraphics[width=1in,height=1.25in,clip,keepaspectratio]{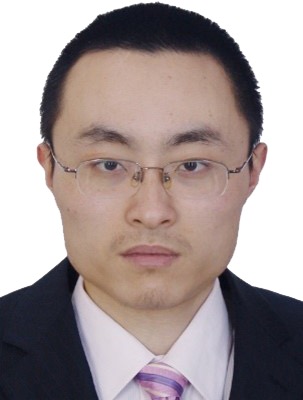}}]{Yudong Zhang (Seinor Member, IEEE)} received a Ph.D. in Signal and Information Processing from Southeast University in 2010. He worked as a postdoc from 2010 to 2012 with Columbia University, USA, and as an Assistant Research Scientist from 2012 to 2013 with the Research Foundation of Mental Hygiene (RFMH), USA. He served as a Full Professor from 2013 to 2017 at Nanjing Normal University. He serves as a Chair Professor from Dec/2017 at the School of Computing and Mathematical Sciences, University of Leicester, UK.
\end{IEEEbiography}

\end{document}